\documentclass[prb,twocolumn]{revtex4}
\usepackage{graphicx}

\begin{document}

\bibliographystyle{revtex}
\title{Flux pinning properties of superconductors with an array of blind holes}

\author{S. Raedts}
\affiliation{Nanoscale Superconductivity and Magnetism Group, Laboratory for Solid State Physics and Magnetism, K.U.Leuven\\
Celestijnenlaan 200 D, B-3001 Leuven, Belgium}

\author{A.~V. Silhanek}
\affiliation{Nanoscale Superconductivity and Magnetism Group,
Laboratory for Solid State Physics and Magnetism, K.U.Leuven\\
Celestijnenlaan 200 D, B-3001 Leuven, Belgium}

\author{M.~J. Van Bael}
\affiliation{Nanoscale Superconductivity and Magnetism Group,
Laboratory for Solid State Physics and Magnetism, K.U.Leuven\\
Celestijnenlaan 200 D, B-3001 Leuven, Belgium}

\author{V.~V. Moshchalkov}
\affiliation{Nanoscale Superconductivity and Magnetism Group,
Laboratory for Solid State Physics and Magnetism, K.U.Leuven\\
Celestijnenlaan 200 D, B-3001 Leuven, Belgium}

\date{\today}

\begin{abstract}

We performed ac-susceptibility measurements to explore the vortex
dynamics and the flux pinning properties of superconducting Pb
films with an array of micro-holes (antidots) and non-fully
perforated holes (blind holes). A lower ac-shielding together with
a smaller extension of the linear regime for the lattice of blind
holes indicates that these centers provide a weaker pinning
potential than antidots. Moreover, we found that the maximum
number of flux quanta trapped by a pinning site, i.e. the
saturation number $n_{s}$, is lower for the blind hole array.

\end{abstract}


\maketitle


\section{Introduction}
The latest advances of lithographic techniques based on electron
beams have allowed to design and tailor artificial pinning centers
in type II superconductors practically at will. In particular, it
has been shown that periodically distributed pinning centers lead
to a strong reduction of the vortex mobility and consequently to a
substantial increase of the critical current when the flux line
lattice is commensurate with the pinning
array.~\cite{daldini,baert95prl,vvm96prb,vvm98prb,mvb01prl,mvb99prb,martin97prl}
So far, most of the work has been devoted to arrays of holes
(antidots)\cite{baert95prl,vvm96prb,vvm98prb} and magnetic
dots.\cite{mvb01prl,mvb99prb,martin97prl} However, much less
attention has been paid to the analysis of blind hole arrays.
Unlike antidots, these non-fully perforated holes have a thin
superconducting bottom layer which allows the trapped flux to
remain as separated single quantum vortices inside the pinning
site. A direct confirmation of this behavior was reported by
Bezryadin \emph{et al}. \cite{Bezryadin96prb} who used vortex
imaging by means of Bitter decoration. On top of that, a blind
hole sample represents a singly-connected system while an antidot
sample is a multiply-connected one. As has been pointed out by
Moshchalkov \emph{et al}. \cite{vvm96prb,vvm98prb} this
topological consideration might also lead to differences in the
irreversible response.

In this work we perform a comparative study of the vortex dynamic
response in type II superconducting Pb films with an array of
blind holes and antidots, by ac-susceptibility $\chi$
measurements.\cite{metlushko99prb,metlushko98euro} We found that
blind holes are less efficient pinning centers than antidots. This
effect manifests itself as a lower ac-shielding and consequently
as a smaller extension of the linear regime. Additionally, we show
that the maximum number of flux quanta, $n_{s}$,
\cite{Mkrtchyan72,Doria,Vinokur} trapped by a blind hole is
systematically lower than for an antidot.

\section{Experimental aspects}
\subsection{Sample preparation}
\begin{figure}[ht]
  \centering
  \includegraphics*[scale=0.65]{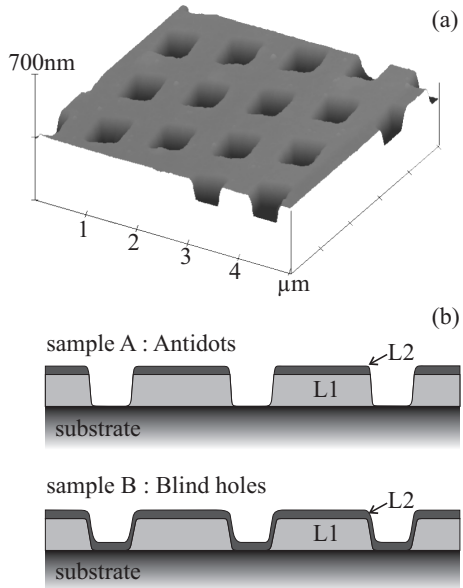}
  \caption{(a) Atomic force micrograph (AFM) of
  a (5~$\times$~5~$)\mu$m$^2$ area of a Pb film with a square array of square
  blind holes. (b) Schematic cross section of the patterned superconducting samples studied in this work,
  a blind hole sample B and an antidot sample A. The two evaporated Pb layers L1 and L2 are indicated.}
  \label{fig:schema}
\end{figure}

The used nanostructured superconducting Pb films were prepared as
follows: first, a superconducting Pb layer is deposited on a
Si/SiO$_2$ substrate covered by a double (PMMA$\backslash$MMA)
resist layer in which a square lattice of square dots is
predefined by electron-beam lithography (Imec vzw). The Pb layer
is deposited in a molecular-beam epitaxy system at a working
pressure of 7~$\times$~$10^{-8}$ Torr. In order to obtain a smooth
Pb film the substrate is cooled by liquid nitrogen (77 K) and the
film is evaporated at a growth rate of 5 \AA/s, controlled by a
quadrupole mass spectrometer. After the evaporation, the remaining
resist is removed by a lift-off procedure using warm aceton. The
double resist layer has an overhanging profile which avoids any
contact of the deposited material on top of the resist dots with
material between the dots. The final result is a Pb film with a
square lattice of square holes. For the protection of the Pb
samples against oxidation a 70 nm-thick Ge capping layer is
finally evaporated on top of the film. In order to grow the
antidot and the blind hole samples simultaneously, we first
deposit a Pb layer (L1) on top of two identical resist dot
patterns. Then, for one of them (sample B in
Fig.~\ref{fig:schema}(b)) we carry out a lift-off procedure
whereas the other (sample A in Fig.~\ref{fig:schema}(b)) remains
unchanged. After that, a second Pb layer (L2) is deposited on top
of both samples. Finally, the resist on sample A is removed by
lift-off. In this way we end up with an antidot sample (sample A)
which has exactly the same thickness as the blind holes (sample B)
and has been grown under identical conditions.

The data presented in this work were obtained from two sets of
blind and antidot samples. Each family has a different total
thickness as determined by low-angle X-ray diffraction. In Table I
we give the thicknesses of the subsequently evaporated Pb layers,
L1 and L2, for the two studied sets of samples.
\begin{table}
\centering \caption{Thicknesses of Pb layers L1 and L2 for the two
sets of studied samples.} \vspace{0.5cm}
\begin{tabular}{|c|c|c|}
  \hline
   & set 1 & set 2 \\
  \hline
  \hspace{0.8cm} L1\hspace{0.8cm}~& \hspace{0.8cm}47.5 nm\hspace{0.8cm} & \hspace{0.8cm}75 nm\hspace{0.8cm} \\
  \hspace{0.8cm} L2\hspace{0.8cm} & \hspace{0.8cm}13.5 nm\hspace{0.8cm} & \hspace{0.8cm}25 nm\hspace{0.8cm} \\
  \hline
  \end{tabular}
  \label{table}
  \end{table}
Fig.~\ref{fig:schema}(a) shows an atomic force microscopy (AFM)
image of a (5~$\times$~5$)\mu$m$^2$ surface area of the blind hole
sample. The lateral size ($b = 0.8~\mu$m) of the holes and the
period of the square array ($d = 1.5~\mu$m) are identical for all
used samples. The periodicity of the square lattice corresponds to
a first matching field of $H_1=\phi_{0}/d^{2}= 9.2 $ Oe. Here
$\phi_{0}$ is the flux quantum.

\subsection{Superconducting properties}

The ac-magnetization measurements were carried out in a commercial
Quantum Design PPMS-system with the ac-field $h$ parallel to the
dc-field $H$ and both applied perpendicular to the sample surface.
This system provides a temperature stability better than 0.5~mK
which is crucial for measurements near the critical temperature.
The ac-amplitude $h$ ranges from 2~mOe to 15~Oe and the frequency
$f$ from 10~Hz to 10~kHz. Since in this range of frequencies we
observe that $\chi$ depends only weakly on $f$, we have chosen the
same frequency $f=3837$ Hz for all measurements presented in this
paper.

\begin{figure}[ht]
  \centering
  \includegraphics[width=90mm]{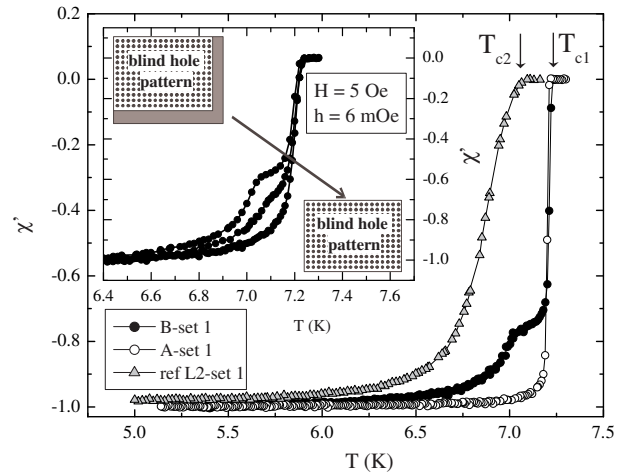}
  \caption{Screening $\chi'$ as function of temperature $T$ for set 1 of Pb films with an array of
  antidots (A, open circles), blind holes (B, filled
  circles) and a reference plain Pb film with the same thickness
  as layer L2 (triangles), with \mbox{$H = 5$ Oe}, \mbox{$f = 3837$ Hz} and \mbox{$h = 6 $ mOe}.
  Inset: $\chi'$ as function of $T$ measured on blind hole
  sample B with the plain Pb contour progressively removed.}
  \label{fig:XvsT61}
\end{figure}

In order to characterize the physical properties of the different
patterned films we first analyze the temperature dependence of the
ac-susceptibility $\chi$~=~$\chi'$~+~$\chi''$. The result of these
measurements for set 1 of samples is shown in the main panel of
Fig.~\ref{fig:XvsT61} at $H = 5$ Oe and $h = 6$ mOe. The data
presented in this figure have been normalized by a factor
corresponding to the maximum screening, such that $\chi' = -1$ at
very low temperatures and fields.
It can be seen that the $\chi'(T)$ curve for the antidot sample A
(open circles) shows a very sharp superconducting transition at
\mbox{$T_{c1} = 7.22$~K}. In contrast to that, the $\chi'(T)$ data
for blind hole sample B (filled circles) first exhibits a sharp
transition at $T_{c1}$ followed by a second broader transition at
\mbox{$T_{c2} = 7.10$ K}, below which it smoothly approaches to
the maximum screening. In Fig.~\ref{fig:XvsT61} we also include
the superconducting transition corresponding to a non-patterned
plain Pb film (triangles) with the same thickness as layer L2 and
evaporated simultaneously with samples A and B. The
superconducting transition of this film coincides with the onset
of the second step on sample B.

The origin of this two-step transition in the blind hole sample
comes from a very narrow Pb border surrounding the blind hole
pattern as a result of the fabrication procedure. Since the
ac-response is mainly given by the border of the sample, a
substantial enhancement of the screening at $T_{c2}$ is expected
when this Pb contour turns to the superconducting state, in
agreement with our observation. In order to test this, we perform
$\chi'(T)$ measurements on a similar sample while progressively
removing the plain Pb contour, as shown in the inset of
Fig.~\ref{fig:XvsT61}. Now, it can be clearly seen that the
transition at $T_{c2}$ first becomes broader and finally
disappears after completely removing the plain Pb border. Although
this undesirable contour may be eventually cut out, it helps to
determine the critical temperature of Pb layer L2 without
preparing an extra plain film. In this case, special care has to
be taken in the normalization process since the total saturation
value at low temperatures results from both, the patterned and the
unpatterned areas.

\section{Results and discussion}
\begin{figure}[ht!]
  \centering
  \includegraphics[width=95mm]{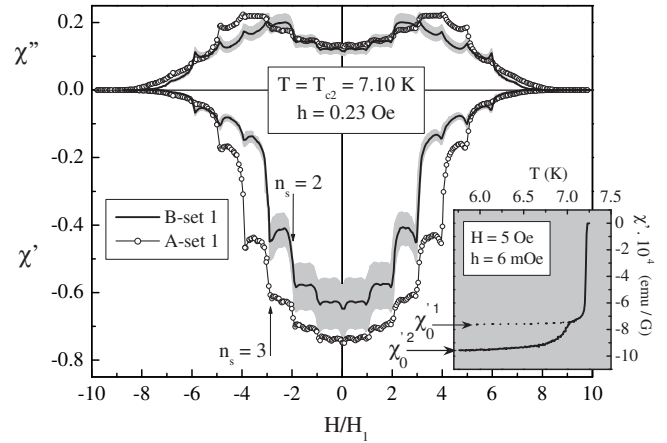}
  \caption{Screening $\chi'$  and dissipation $\chi''$ for films of set 1 with an array of
  antidots (open circles) and blind holes (thick solid line) as function of $H/H_{1}$ for \mbox{$T=T_{c2}=7.10$
  K} and \mbox{$h = 0.23$ Oe}. The inset shows the $\chi'(T)$ transition for blind hole sample B, indicating the two possible saturation values used in the
  normalization of the signal $\chi'$.}
  \label{fig:XvsH_61nm}
\end{figure}

Let us now compare the flux pinning properties of the blind hole
array with those obtained for the antidot array. To that end we
have carried out measurements of the ac-response in samples A and
B as a function of dc-field under isothermal conditions and fixed
ac-excitations. This is shown in the main panel of
Fig.~\ref{fig:XvsH_61nm} for $h = 0.23$~Oe, \mbox{$T = T_{c2} =
7.10$ K} and $f = 3837$ Hz. In agreement with previous
reports,\cite{vvm96prb,vvm98prb,baert95prl,SilhanekPRBns} the
antidot sample A (open symbols) exhibits clear periodic matching
features at integer and rational multiples of the first matching
field $H_1$. As we have discussed in an earlier
work,\cite{SilhanekPRBns} two different regimes can be
distinguished in this curve. At low fields $H < H_3$, a
multi-quanta vortex state exists and matching features appear as
small steps of the screening $\chi'$. For fields $H > H_3$ the
filled pinning sites become repulsive centers and entering
vortices locate in the interstitial positions. In this regime,
vortex-vortex interaction leads to highly stable vortex
configurations at $H_n$ thus resulting in local enhancements of
the screening $\chi'(H)$. We have also shown\cite{SilhanekPRBns}
that the sharp reduction in the screening at $H_4$ can be
attributed to the higher sensitivity of the ac-susceptibility in
that particular range of field penetration.

As we have pointed out above, the analysis of the blind hole
sample is a more subtle procedure since the signal normalization
can be derived either from the saturation value corresponding to
the first or the second transition. For example, data taken at
$T>T_{c2}$, where only the patterned film contributes to the
signal, should be normalized using the saturation value obtained
by extrapolating the first transition ($\chi'^1_0$), as shown with
a dotted line in the inset of Fig.~\ref{fig:XvsH_61nm}. A
different normalization value could be obtained due to proximity
effects which lead to a larger effective sample size and
consequently to a higher saturation. However, no substantial
change of $T_{c}$ has been detected, suggesting that the proximity
effect is not relevant. In any case, the correct normalization
value will lay between the two extreme values $\chi'^1_0$ and
$\chi'^2_0$, indicated by black arrows in the inset of Fig.
\ref{fig:XvsH_61nm}. The result of this normalization procedure is
shown as a continuous curve in the main panel of Fig.
\ref{fig:XvsH_61nm}, whereas the extremes obtained by normalizing
with $\chi'^1_0$ and $\chi'^2_0$ are shown as a gray painted area.
The saturation value $\chi'_{0}$ can be also estimated as
$\chi'_{0}=\frac{V}{4\pi (1-\nu)}$ where $V$[cm$^{3}$] is the
volume of the sample and $\nu$ the demagnetization factor.
\cite{clem-sanchez,poole} For this particular sample with lateral
dimensions $w_{1}$ and $w_{2}$ and thickness $\delta$,
$V=w_{1}\times w_{2}\times \delta \approx
4.7\times10^{-7}$cm$^{3}$ and $1-\nu\sim
\frac{\delta}{w_{1}}+\frac{\delta}{w_{2}}\sim 3.8\times10^{-5}$,
so \mbox{$\chi_{0}\sim 9.8\times10^{-4}$ emu/G} which is very
close to the experimental value \mbox{$\chi'^2_0 = 9.7\times
10^{-4}$ emu/G}. Regardless the chosen normalization, we can
clearly see that commensurability features are also present in the
blind hole sample.

A direct comparison of the $\chi(H)$ curves for samples A and B
allows us to identify two clear differences. First, the overall
screening is lower for sample B, indicating that blind holes
provide a less efficient pinning. This effect can be intuitively
understood by considering the two extreme limits of very shallow
blind holes (plain film) where only intrinsic defects pin the
vortices, and very deep blind holes (antidots) with a much
stronger pinning force. Within this picture, it is expected that
the effective pinning force grows continuously as the thickness of
the bottom layer decreases. The second point to consider is that
$n_{s}=2$ for blind holes whereas $n_{s}=3$ for antidots (see
black arrows in the main panel of Fig.~\ref{fig:XvsH_61nm}). The
same difference in $n_{s}$ was found by performing
dc-magnetization measurements on the same set of samples. This
result is consistent with previous Bitter decoration
experiments\cite{Bezryadin96prb} showing that the difference
between the saturation number of blind holes and antidots does not
exceed one.
\begin{figure}[ht]
  \centering
  \includegraphics*[scale=1]{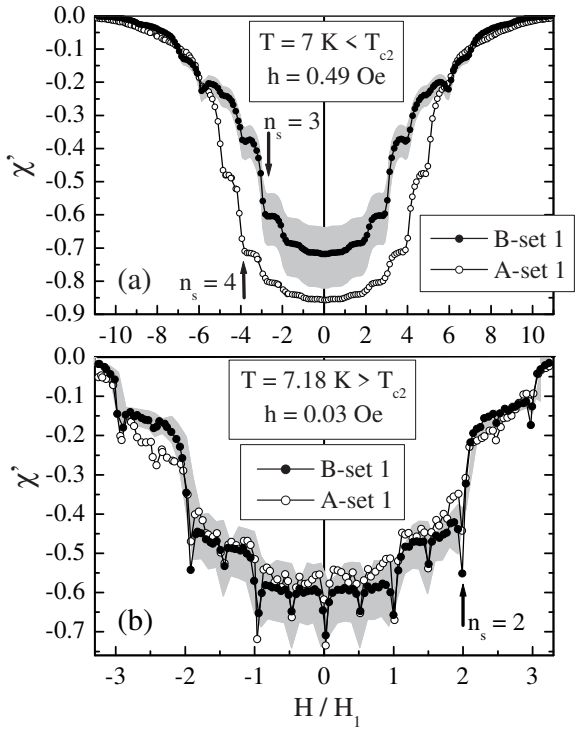}
  \caption{Screening $\chi'$ as function of $H/H_{1}$ for Pb films of
  set 1 with an array of blind holes (filled symbols) and antidots (open symbols)
  with (a) $T=7 K < T_{c2}$ and $h = 0.49$ Oe and (b) $T = 7.18 K > T_{c2}$ and $h = 0.03$ Oe.}
  \label{fig:XvsH 61 nm diff T}
\end{figure}

The origin of these differences can be attributed to the pinning
nature of blind holes and antidots. Indeed, the interaction of a
flux line with a blind hole substantially differs from the more
widely investigated vortex-antidot interaction. In both cases, the
normal/superconductor boundary imposes a condition to the
supercurrents to flow parallel to the boundary of the hole. This
effect can be modelled by introducing an image antivortex inside
the hole which interacts attractively with the flux
line.\cite{Buzdin} For the antidots, this attractive force acts
along the total length of the flux line, whereas for blind holes
we expect, as a first approximation, a smaller force proportional
to the depth of the hole. This scenario becomes more complicated
when considering the interaction of a flux line with an occupied
blind hole. In this case, whereas flux quanta trapped by an
antidot consist of supercurrents flowing around the hole, flux
quanta pinned by blind holes remain as separated single-quanta
flux lines with a well defined core. Now an external vortex
outside of the blind hole would simultaneously feel attraction due
to the image antivortex and repulsion due to the trapped vortex.
Besides that, the stray field produced by vortices inside the
blind holes can not spread out freely in space since it has to be
screened by the inner edges of the hole, this leads to an extra
term in the interaction. For higher fillings, trapped flux lines
are able to rearrange inside the blind hole, a degree of freedom
absent in antidots. The repulsive interaction between these
single-quanta vortices might explain the origin of the lower
saturation number observed for the blind hole sample.

Let's now move on to the analysis of the ac-response for
temperatures above and below the critical temperature, $T_{c2}$ of
the bottom layer. For $T<T_{c2}$, as expected, we observe the same
different flux pinning properties for blind holes and antidots, as
is shown in Fig.~\ref{fig:XvsH 61 nm diff T}(a) for \mbox{$T = 7$
K}. For $T>T_{c2}$, an isolated plain Pb film with the same
thickness as layer L2 is in the normal state (see
Fig.~\ref{fig:XvsT61}). Although this film L2 forms the bottom
layer of the blind holes, in this case it is not isolated but
rather surrounded by the superconducting Pb bilayer which may
induce superconductivity. Therefore, in this specific temperature
region we expect that the pinning behavior of blind holes
asymptotically approaches that of the antidots. This is indeed
confirmed by the data shown in Fig.~\ref{fig:XvsH 61 nm diff T}(b)
for the same set of samples at \mbox{$T=7.18 $~K}. The most
obvious feature of this figure is the similarity between the
ac-response of both samples, i.e. similar ac-shielding and the
same saturation number.
\begin{figure}[ht]
  \centering
  \includegraphics[width=90mm]{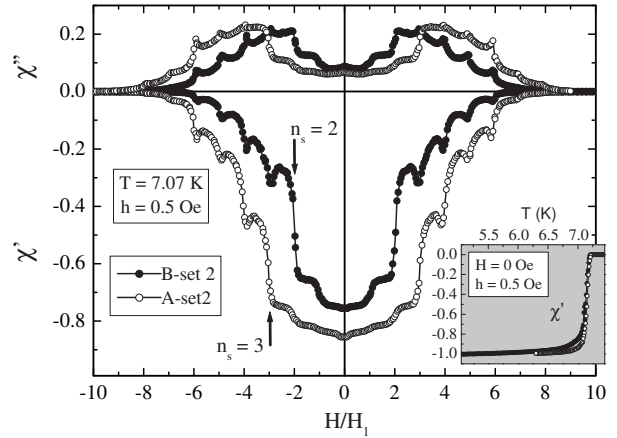}
  \caption{Screening $\chi'$ and dissipation $\chi''$ as function of $H/H_{1}$, for Pb films of set 2 with an array of
  antidots (open circles) and blind holes (filled circles) for $T = 7.07$ K, $f = 3837$ Hz and $h = 0.5$ Oe.
  The inset shows the temperature dependence of the normalized screening $\chi'$ for the samples A and B.}
  \label{fig:XvsH_100nm}
\end{figure}
All the observations reported for set 1 of samples were also
reproduced for set 2 of samples. These results are shown in
Fig.~\ref{fig:XvsH_100nm}. In this case, sample A and B have the
same $T_{c}=7.22$ K, as is shown in the inset of
Fig.~\ref{fig:XvsH_100nm}.

\begin{figure}[h!]
  \centering
  \includegraphics*[scale=1]{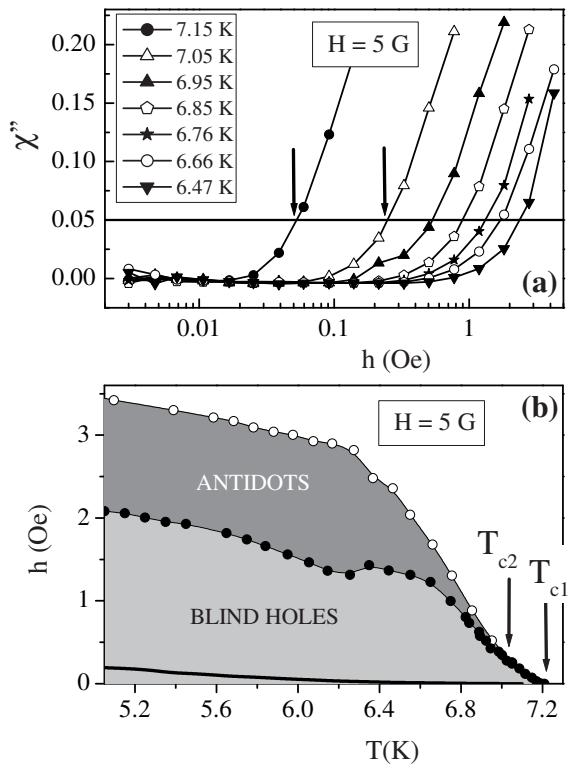}
  \caption{(a) Dissipation $\chi''$ as function of the ac-field $h$ for an array of
  antidots at several $T$, $f = 3837$ Hz and $H = 5$ Oe. Arrows indicate the onset of the non-linear response
  according to the chosen criterium $\chi''=0.05$ (horizontal line). (b) Phase boundary of the linear regime for samples A and B of set 1,
  for $H = 5$ Oe and $f = 3837$ Hz. This boundary is obtained using a dissipation criterium $\chi'' = 0.05$
  as shown in (a) for antidot sample A. The continuous line indicates the boundary of the linear regime
  for a reference non-patterned Pb film with the same thickness as layer L2.}
  \label{fig:boundaryLR}
\end{figure}

An alternative way to investigate the pinning properties of blind
holes and antidots is to analyze the different ac vortex dynamic
regimes.\cite{pasquini99prb,SilhanekEPJB} For very low ac-drives,
all vortices oscillate inside the corresponding individual pinning
potentials. This so-called linear regime is characterized by an
$h$-independent screening together with a very low
dissipation.\cite{pasquini97physicaC,Silhanek03prb} As the
ac-drive is increased, vortices eventually overcome the pinning
well switching to a more dissipative regime with an $h$-dependent
screening. The boundary between these two regimes is mainly
determined by the strength of the pinning centers. Consequently,
the stronger the pinning, the larger the extension of the linear
regime. Experimentally, a reliable criterium to determine the
onset of non-linearity is given by a dissipation \mbox{$\chi''(h)
= 0.05$} as is shown in Fig.~\ref{fig:boundaryLR}(a) for sample A
of set 1 at several temperatures. Performing this procedure for
samples A and B, we can compare the dynamic diagrams $h(T)$ of
antidot and blind hole samples (see Fig.~\ref{fig:boundaryLR}(b)).
Most obvious in Fig.~\ref{fig:boundaryLR}(b) is the smaller
extension of the linear regime for the blind hole sample B. This
is a clear indication that the blind hole array produces a weaker
pinning potential, in agreement with our previous observations. In
addition, for temperatures $T>T_{c2}$, \emph{the two boundaries
collapse on a single line}. This result is consistent with the
fact that for $T>T_{c2}$ the thin layer at the bottom of the blind
holes approaches to a normal metal, thus turning to the behavior
of the antidot sample. Fig.~\ref{fig:boundaryLR}(b) also includes
the dynamic diagram $h(T)$ for a reference film with the same
thickness as layer L2. As expected, the very low effective pinning
of the plain film results in a substantial smaller extension of
the linear regime in comparison with the patterned samples A and
B.

It is important to stress that there is also a difference in the
depinning process of vortices trapped by antidots and blind holes.
On one hand, single-quanta vortices trapped by the blind holes are
able to depin one by one. On the other hand, as has been pointed
out by Priour and Fertig,\cite{Priour-Fertig} in the case of
multiquanta vortices (without rigid core) trapped by antidots, the
driving current elongates the vortex core which can eventually
reach the neighbor pinning site thus allowing the vortex to hop
from site to site. All these considerations should be taken into
account in order to theoretically analyze the pinning properties
of blind holes.

\section{Conclusion}
We have used ac-susceptibility to perform a comparative study of
the flux pinning properties of an array of antidots and blind
holes. We show that antidots are more efficient pinning centers
than blind holes where the superconducting film is not fully
perforated. Consequently, a reduced screening for the blind hole
system is observed. Therefore, the strength of the pinning
potential can be gradually tuned by varying the depth of blind
holes. On top of that, the saturation number $n_{s}$, defined as
the maximum number of flux quanta that a pinning site can hold, is
higher for antidots than for blind holes, in agreement with
previous reports. The linear regime, in which vortices oscillate
inside the pinning potential, has a smaller extension for the
blind hole sample, indicating that blind holes provide a weaker
pinning potential. Finally, we discussed the ac-response for
temperatures above the critical temperature of the bottom layer
and found that the pinning behavior of blind holes approaches the
behavior of antidots.

\section{Acknowledgements}
This work was supported by the Fund for Scientific
Research-Flanders (FWO-Vlaanderen), the Belgian Inter-University
Attraction Poles (IUAP), the Research Fund K.U.Leuven GOA/2004/02
and by the European ESF VORTEX programs. MJVB is a postdoctoral
Research Fellow of the FWO. We thank J. Van de Vondel and C. de
Souza Silva for helpful discussions.

\end{document}